\documentclass[fleqn,10pt]{wlscirep}
\usepackage{amssymb, amsmath, amsfonts, amsthm, graphicx, subfig, color,soul,array,booktabs,lineno,prettyref,cite,hyperref}
\usepackage[section]{placeins}

\title{Correlates of severity of disease in \textit{Macaca mulatta} infected with \textit{Plasmodium cynomolgi}}
\begin{document}
\author[1+]{Yi H. Yan}
\author[2+]{Diego M. Moncada}
\author[1]{Elizabeth D. Trippe}
\author[1,3,*]{Juan B. Gutierrez}

\affil[1]{University of Georgia, Institute of Bioinformatics, Athens, 30602, USA}
\affil[2]{Universidad del Quindio (Colombia)}
\affil[3]{University of Georgia, Department of Mathematics, Athens, 30602, USA}

\affil[*]{mailto:jgutierr@uga.edu}
\affil[+]{These Authors Contributed To This Work Equally}
\begin{abstract}

Characterization of host responses associated with severe malaria through an integrative approach is necessary to understand the dynamics of a \textit{Plasmodium cynomolgi} infection. In this study, we conducted temporal immune profiling, cytokine profiling and transcriptomic analysis of five \textit{Macaca mulatta} infected with \textit{P. cynomolgi}. This experiment resulted in two severe infections, and two mild infections. Our analysis reveals that differential transcriptional up-regulation of genes linked with response to pathogen-associated molecular pattern (PAMP) and pro-inflammatory cytokines is characteristic of hosts experiencing severe malaria. Furthermore, our analysis discovered associations of transcriptional differential regulation unique to severe hosts with specific cellular and cytokine responses. The combined data provide a molecular and cellular basis for the development of severe malaria during \textit{P. cynomolgi} infection.
\end{abstract}
\date{\today}
\maketitle 

\section{Introduction}

	Malaria remains a public health challenge, responsible for approximately 400,000 (236,000–635,000) death in 2015 \cite{world2016world}. Out of the five human \textit{Plasmodium} species capable of causing malaria, \textit{Plasmodium falciparum} and \textit{Plasmodium vivax} account for the majority of human malaria infections. \textit{P. falciparum} is responsible for the majority of malaria-related mortality and is most prevalent in sub-Saharan Africa \cite{world2016world}. Although \textit{P. vivax} infection does not typically result in mortality, this parasite causes substantial morbidity outside of sub-Saharan Africa and exhibits the propensity to cause severe disease. The mechanisms that underly vivax malaria pathogenesis, however, remain poorly understood partially due to experimental constraints such as the lack of an \textit{in vitro} culture system and a rodent model of \textit{P. vivax} infection \cite{mueller2009key,galinski2013plasmodium,noulin20131912}. 
%
%

\textit{Plasmodium cynomolgi} is a non-human primate parasite that infects old world monkeys and is capable of recapitulating clinical and histopathological findings of vivax malaria patients\cite{deye2012use,galinski2008plasmodium,tachibana2012plasmodium}. It is both genetically and physiologically similar to \textit{P. vivax} \cite{tachibana2012plasmodium,waters1993evolutionary}. For instance, both parasites exhibit 48-h erythrocytic cycle during blood stage infection \cite{coatneyprimate}, preferential infection of reticulocytes \cite{warren1966biology} and form hypnozoites, which are dormant, liver stage forms that can activate and cause relapse infections \cite{krotoski1982observations}. Due to the difficulty of studying \textit{P. vivax} pathogenesis, \textit{P. cynomolgi} infection of rhesus monkeys (\textit{Macaca mulatta} ) has being used to better understand hypnozoite-caused relapse \cite{cogswell1992hypnozoite}.

%
%
%
%

Host clearance of malaria parasites without complication requires a concerted effort between inflammatory and anti-inflammatory cytokines; their balance and timing are critical in determining clinical outcomes \cite{gonccalves2014parasite}. The association between cytokine, transcriptomic and immune response, and clinical outcomes has been extensively studied in \textit{P. falciparum} infections \cite{prakash2006clusters,tran2016transcriptomic,torres2014relationship,wykes2014malaria,chavale2012enhanced}, in comparison, much less is known for \textit{P. vivax} infections.

With the aim to better characterizing \textit{P. vivax} infection in humans, a time series experiment where four \textit{Macaca mulatta} were infected with \textit{P. cynomolgi} was conducted as a part of the Malaria Pathogen Host Interaction Center (MaHPIC) project \cite{joyner2016plasmodium}. This experiment captured host transcriptomic, cellular and cytokine response to \textit{P. cynomolgi} infection. The subjects within this study responded to the infection in different manners and resulted in two cases of severe malaria, and two cases of mild malaria. The presence of both severe malaria and mild malaria provided us the opportunity to characterize host responses associated with clinical outcomes.  

Here we present the combined analysis of cellular, transcriptome and cytokine data collected during this experiment with an emphasis on comparing host responses during severe malaria and mild malaria. Our analysis shows associations of severe malaria with differential up-regulation of pro-inflammatory signals independent of adaptive immune response, whereas moderate malaria is associated with transcriptomic up-regulation of complement and heme metabolism related genes. These distinct characteristics demonstrate that \textit{P. cynomolgi} induced severe malaria is strongly associated with up-regulation of inflammatory signaling through transcriptional regulation.

\section{Results}
\subsection{Study Design}
\begin{figure}[!ht]
	\centering
	\includegraphics[width = \textwidth]{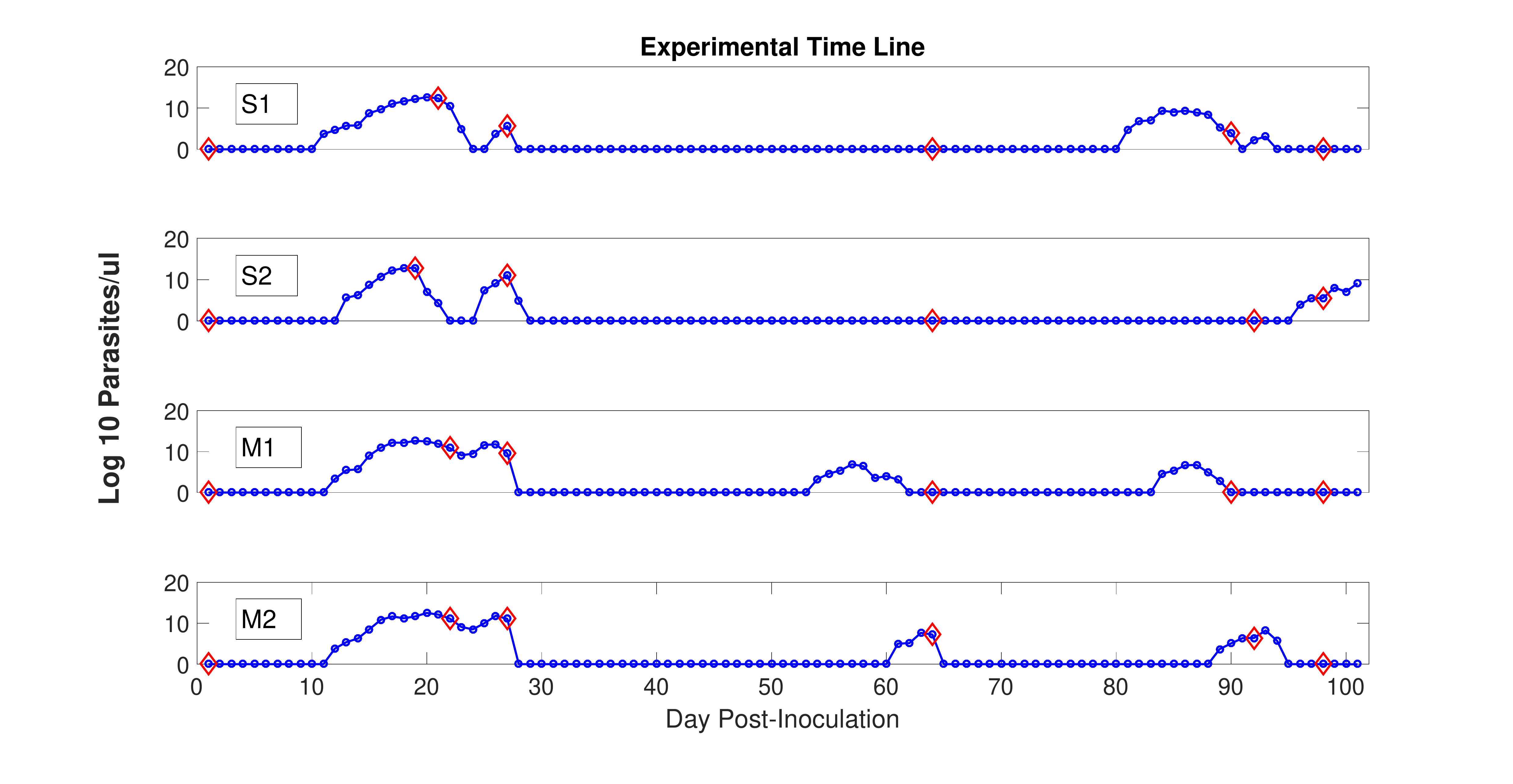}
	\caption{Parasite count over the entire experiment. Transcriptome, cytokine, and immune profiling were conducted during the six time points marked with red diamonds. Subject S1 and S2 experienced severe malaria symptom. Subject M1 and M2 experienced mild malaria symptom.}
	\label{fig:Experimental Set Up}
\end{figure}

The experimental design is shown in Figure \ref{fig:Experimental Set Up}. Transcriptomic, cellular and cytokine data were measured from peripheral blood samples collected during six time points. Time point 1 (TP1) corresponds to the pre-infection baseline for each subject. Time point 2 (TP2) corresponds to the acute primary infection when parasite counts peaked. At time point 2, two subjects, S1 and S2 developed signs of severe malaria including anemia and thrombocytopenia, which required each animal to be sub-curatively treated to prevent possible severe complications. The other two subjects M1 and M2 experienced mild disease manifestations and did not receive sub-curative treatment. All animals received curative blood-stage treatment with artemether on Day 28 to ensure that subsequent parasitemias were due to relapses and not recrudescing parasitemias. A detailed description of the experimental set-up can be found in the publication by Joyner \textit{et al.} \cite{joyner2016plasmodium}.

\subsection{Differential Gene Regulation during Primary Infection}
\begin{figure}[!ht]
	\centering
	\includegraphics[height = 0.6\textheight]{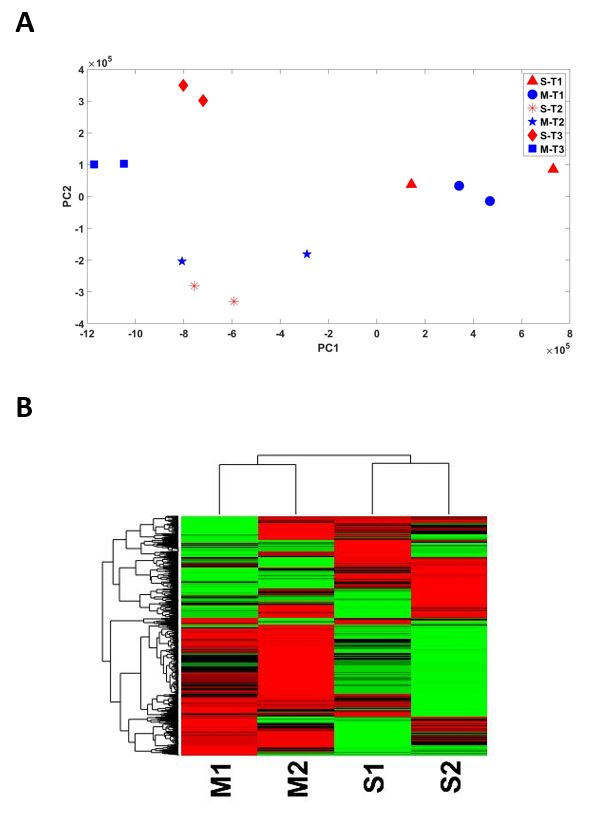}
	\caption{Clinical classes were defined as in Figure \ref{fig:Experimental Set Up}. (A) Principal component analysis of the combined -omic data (Transcriptome, cytokine and immune cell counts). (B) An unsupervised hierarchical heat map of the top 50$\%$ most variable genes at time point 2. Subjects experiencing severe malaria symptoms are labeled S1 and S2, subjects experiencing mild malaria symptoms are labeled M1 and M2.}
	\label{fig:PCA}
\end{figure}

To characterize the differences of transcriptomic responses induced by \textit{P. cynomolgi} infection between the subjects experiencing severe malaria (severe hosts) and the ones experiencing mild malaria (mild hosts), we conducted RNA-Seq analysis for each clinical group between TP1 and TP2 using the DESeq algorithm \cite{anders2010differential}. Furthermore, between-group transcriptome analysis was also carried out for TP2. The transcriptome profiles differ the most from the healthy baseline (TP1) at TP2 and TP3 shown by principal component analysis (Figure \ref{fig:PCA}). Additionally, the transcriptome profiles of the severe hosts were more similar to each other than to the mild hosts at TP2 as shown by both principal component analysis and hierarchical clustering using the top 50$\%$ most variably expressed genes (Figure \ref{fig:PCA}).

%
%

Mild hosts showed similar degrees of transcriptome perturbation during the onset of primary infection (698 DEGs using an absolute fold-change threshold of $>$ 1.5 and q-value $<$ 0.05) relative to severe hosts (511 DEGs). 148 genes were identified by both analyses. At TP2, 1337 genes were differentially expressed between the mild hosts and severe hosts. 380 of these DEGs were temporally differentially expressed in severe hosts between TP1 and TP2. 220 DEGs were temporally differentially expressed in mild hosts between TP1 and TP2.

\begin{figure}[!ht]
	\centering
	\includegraphics[width = \textwidth]{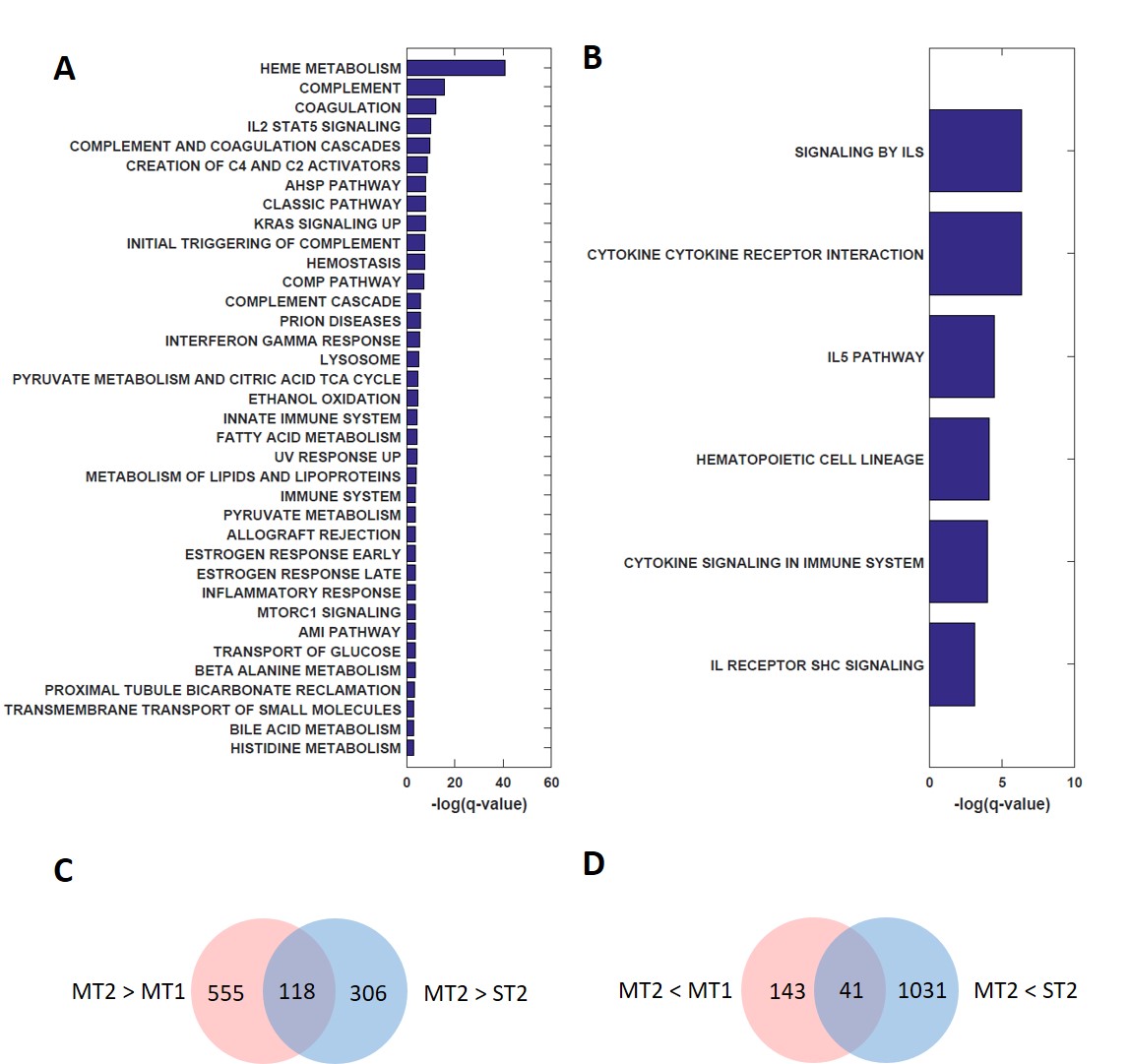}
	\caption{(A) Enrichment analysis using DUGs-M with q-value $< 0.05$. (B)  Enrichment analysis using DDGs-M with q-value $< 0.05$. (C) and (D) Venn diagrams showing the identification of DUGs-M and DDGs-M. $MT2 > MT1$ refers to genes within mild hosts that are up-regulated at TP 2 in comparison to TP 1. $MT2 > ST2$ refers to genes that are up-regulated in mild hosts in comparison to severe hosts at TP2. $MT2 < MT1$ refers to genes within mild hosts that are down-regulated at TP 2 in comparison to TP 1. $MT2 < ST2$ refers to genes that are down-regulated in mild hosts in comparison to severe hosts at TP2. }
	\label{fig:M_Enrichment}
\end{figure}

To identify differentially regulated transcriptome responses to \textit{P. cynomolgi} between severe hosts and mild hosts, DEGs were classified into four groups: differentially up-regulated genes in severe hosts (DUGs-S), differentially up-regulated genes in mild hosts (DUGs-M), differentially down-regulated genes in severe hosts (DDGs-S) and differentially down-regulated genes in mild hosts (DDGs-M). Genes were classified into each of these groups based on their between group differential expression at TP2 and their with-in group differential expression between TP1 and TP2. The specific classification scheme is shown in Figure \ref{fig:S_Enrichment} and Figure \ref{fig:M_Enrichment}.

%
%

\subsection{Biological Pathways Perturbed During Primary Infection}
\begin{figure}[!ht]
	\centering
	\includegraphics[width = \textwidth]{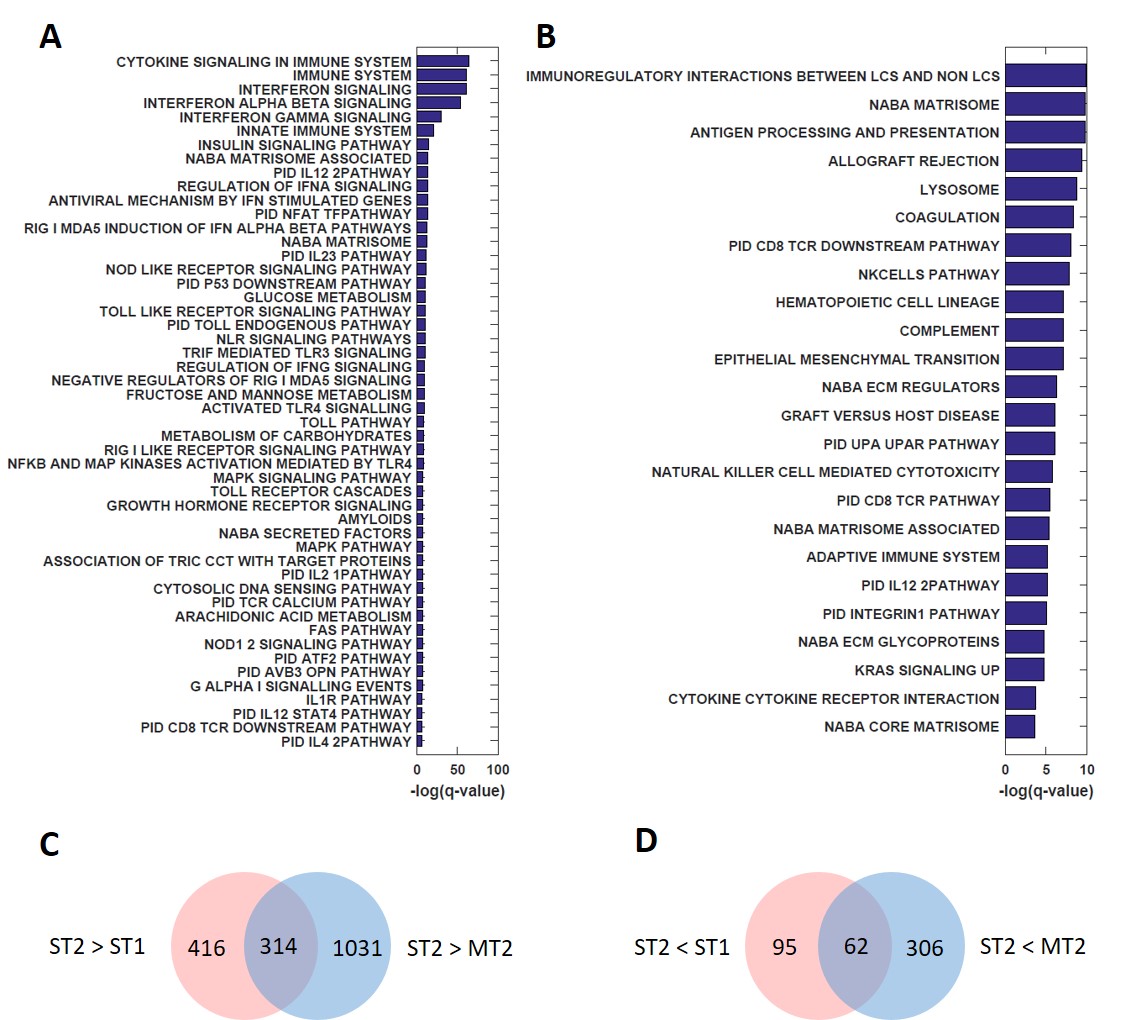}
	\caption{(A) Enrichment analysis using DUGs-S with q-value $< 0.05$. (B)  Enrichment analysis using DDGs-S with q-value $< 0.05$. (C) and (D) Venn diagrams showing the identification of DUGs-S and DDGs-S. $ST2 > ST1$ refers to genes within severe hosts that are up-regulated at TP 2 in comparison to TP 1. $ST2 > MT2$ refers to genes that are up-regulated in severe hosts in comparison to mild hosts at TP2. $ST2 < ST1$ refers to genes within severe hosts that are down-regulated at TP 2 in comparison to TP 1. $ST2 < MT2$ refers to genes that are down-regulated in severe hosts in comparison to mild hosts at TP2. }
	\label{fig:S_Enrichment}
\end{figure}

To identify the biological pathways associated with each of the four groups of differentially regulated genes, we used MySigDB to conduct enrichment analysis \cite{liberzon2011molecular}. Using a q-value cut-off of 0.05, DUGs-S are enriched in 205 pathways with the most significantly enriched pathway being cytokine signaling in the immune system. DDGs-S are enriched in only 24 pathways (Figure \ref{fig:S_Enrichment}) with the most significantly enriched pathway being immunoregulatory interactions between lymphocyte cells (LCS) and non-lymphocyte cells (Non - LCS). DUGs-M is enriched in 51 pathways with the most significantly enriched pathway being heme metabolism, and DDGs-M are enriched in 6 pathways. The number of enriched pathways within each gene set is proportional to the gene set size. Both DUGs-S and DUGs-M show enrichment in innate immune-related response and interferon gamma related response. However, complement and coagulation-related genes are differentially up-regulated in the mild hosts and differentially down-regulated in the severe hosts.  

%
%

\subsection{Differentially Regulated Genes Demonstrate Distinct Correlation Pattern with Cell Populations and Cytokine Abundance}

\begin{figure}[!ht]
	\centering
	\includegraphics[width = \textwidth]{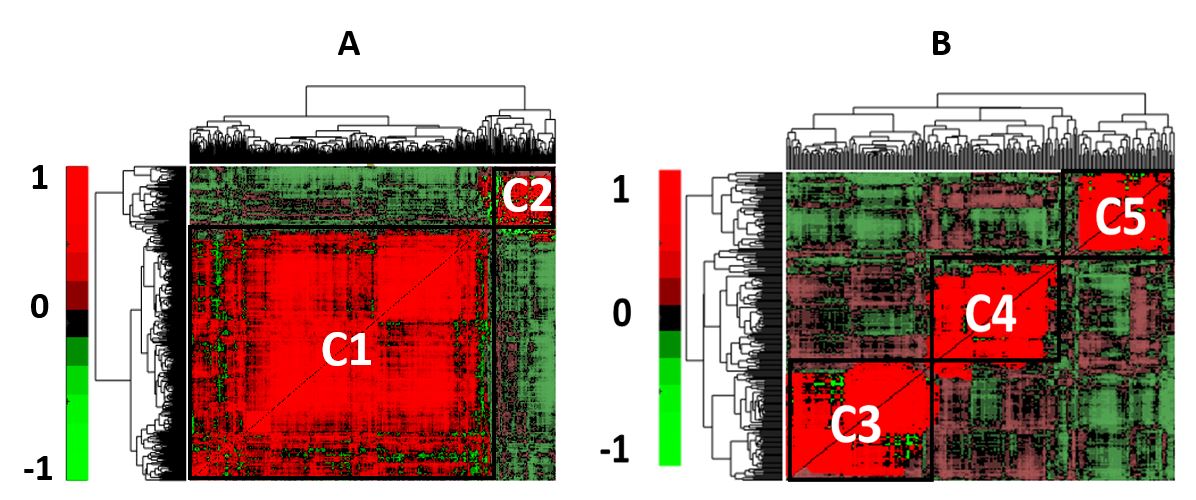}
	\caption{(A) An unsupervised hierarchical heat map of the pairwise Spearman's correlation among DUGs-S, cell populations, and cytokine abundances. (B) Heat map of the pairwise Spearman's correlation among DUGs-M, cell populations, and cytokine abundances.}
	\label{fig:GeneCorr3}
\end{figure}

To identify the similarities among temporal profiles of DUGs-S, DUGs-M, cytokine abundances and cell populations, we constructed two correlation matrices and identified 5 clusters (Figure \ref{fig:GeneCorr3}). DUGs-S and DUGs-M exhibited distinct correlation patterns with respect to cytokine abundances and cell populations. Two clusters were identified for DUGs-S, all but 11 genes were clustered into C1 and the rest into C2. C1 contained the majority of cytokine abundances along with a few cell populations such as the percentage of CD14 CD16 monocytes out of total monocyte, the percentage of granulocytes out of total lymphocytes and parasite concentration. Furthermore, we performed enrichment analysis of transcriptional factors for C1 and identified a list of enriched transcriptional factor targets. The top 3 transcriptional factor targets enriched are ISRE, IRF7, and STAT5. Incidentally, IRF7 was also observed to be transcriptionally differentially up-regulated in severe hosts. Unlike C1, C2 contained most of the adaptive and innate immune cell populations but very few genes. In fact, the majority ( $> 60 \%$) of the correlations between the genes in C1 and the adaptive immune cell populations in C2 are negative. 

Three clusters were identified for DUGs-M. C3 contained 37 genes and the majority of the pro-inflammatory cytokines along with reticulocyte population and parasite concentration. Enrichment analysis of C3 reveals that it is enriched in heme metabolism related genes. C4, on the other hand, contained the rest of DUGs-M along with caspase 3 CD 8 and CD 4 T cell populations, monocyte and granulocyte populations. C4 is enriched in complement and coagulation-related genes. C5 is composed entirely of adaptive immune cell types such as B cell and traditional T cell populations. Transcription factor analysis of C3 and C4 did not reveal any enrichment of transcription factor targets. Despite the fact that both heme metabolism and complement-related genes are differentially up-regulated at TP2 in mild hosts, their expression patterns are associated with different cell populations and cytokine abundances. Furthermore, their temporal expressions differ over the entire experiment, in contrast to the uniform temporal expression patterns of DUGs-S.

\subsection{Cytokine Perturbation during Primary Infection}

\begin{figure}[!ht]
	\centering
	\includegraphics[width =0.8 \textwidth]{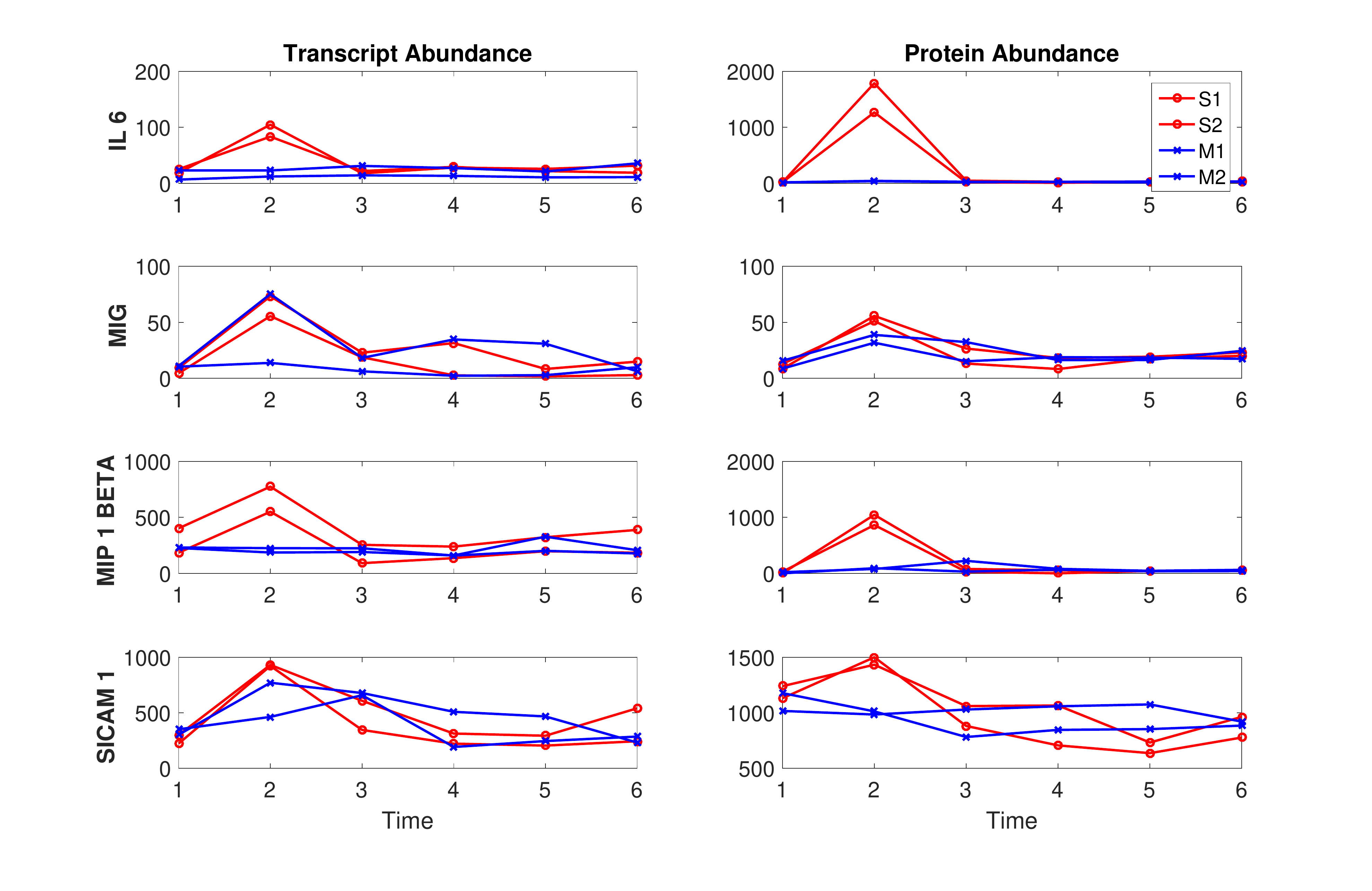}
	\caption{Time series data of significantly different cytokines between severe hosts and mild hosts at time point 2. The first column indicates transcript abundance and the second column indicates protein abundance.}
	\label{fig:TS}
\end{figure}

To characterize the differences of cytokine responses induced by \textit{P. cynomolgi} infection between the clinical groups, two-way ANOVA was used to identify significantly changed cytokines. 4 cytokines out of all 44 cytokines measured displayed significantly different abundances between the two clinical groups. All 4 cytokines have a higher abundance in severe hosts at TP2 (Figure \ref{fig:TS}).

\begin{figure}[!ht]
	\centering

\includegraphics[width = 0.6\textwidth]{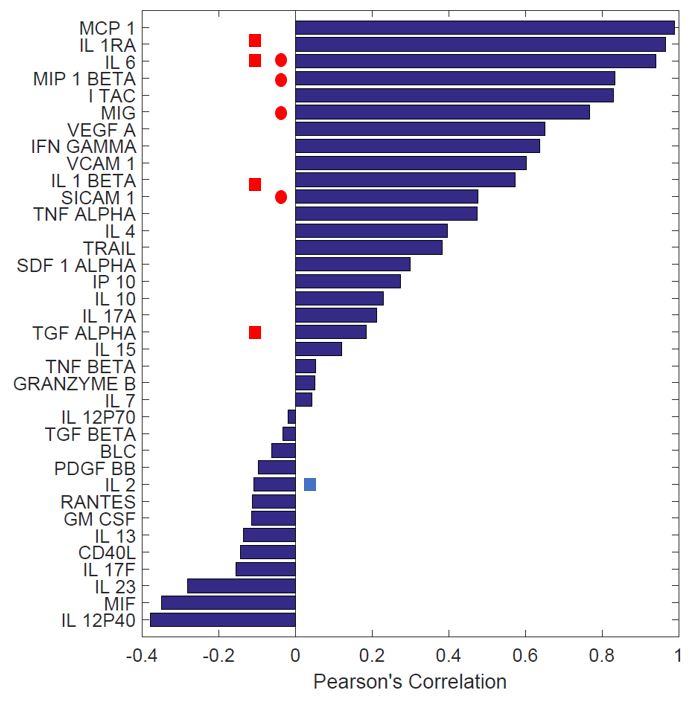}
	\caption{Pearson's correlation between cytokine concentrations and their transcript abundances. A red circle indicates protein level differential up-regulation in severe hosts. A red square indicates transcriptional differential up-regulation in severe hosts. A blue square indicates transcriptional differential up-regulation in mild hosts.}
	\label{fig:GeneCorr2}
\end{figure}

To determine whether the differences in cytokine abundances can be attributed to transcription levels, Pearson's correlation coefficients between the time series of each of the cytokines and their respective transcripts were calculated (Figure \ref{fig:TS}, \ref{fig:GeneCorr2}). The correlations of IL-6, MIG, MIP-1 and SICAM1 (soluble ICAM1) to their transcripts are 0.94, 0.77, 0.83 and 0.48 respectively. The correlation between SICAM1 and its transcript within the severe hosts is 0.59, which is significantly higher than that of the mild hosts (correlation of 0.17). Additionally, MIG, SICAM1, and MIP-1 only differ in cytokine abundances but not transcriptional abundances. The transcription regulators of IL-6, MIG, MIP-1 and SICAM1 were identified using the TRRUST database \cite{han2015trrust} and analyzed for differential up-regulation in severe hosts. One transcription factor of IL-6, JUN, and one transcription factor of ICAM1, TWIST2, were identified as differentially up-regulated in severe hosts at TP2. 

%
%

%
%

Interestingly, 13 cytokines exhibited negative correlation between their protein abundances and transcriptional abundances. One of these 13 cytokines, IL-2 is transcriptionally differentially up-regulated in mild hosts at TP2. Additionally, all the genes belonging to DUGs-S showed positive correlations to their protein abundances. MCP1 and IL1RA exhibit the highest correlation between transcript abundances and protein abundances. Both IL1-beta and IL1RA showed differential up-regulation in severe hosts at the transcriptional level, however, their transcript-protein correlation differs. IL-1-beta showed a 5 fold transcriptional change which is associated with a 4 fold increase protein abundance. IL-1RA showed 7 fold increase in transcripts but $\sim 8000$ fold change in protein abundance. The ratio of IL1 to IL1RA is lower in severe hosts (0.0045) in comparison to mild hosts (0.04). 

\section{Discussion}

\begin{figure}[!ht]
	\centering
	\includegraphics[width = 0.7\textwidth]{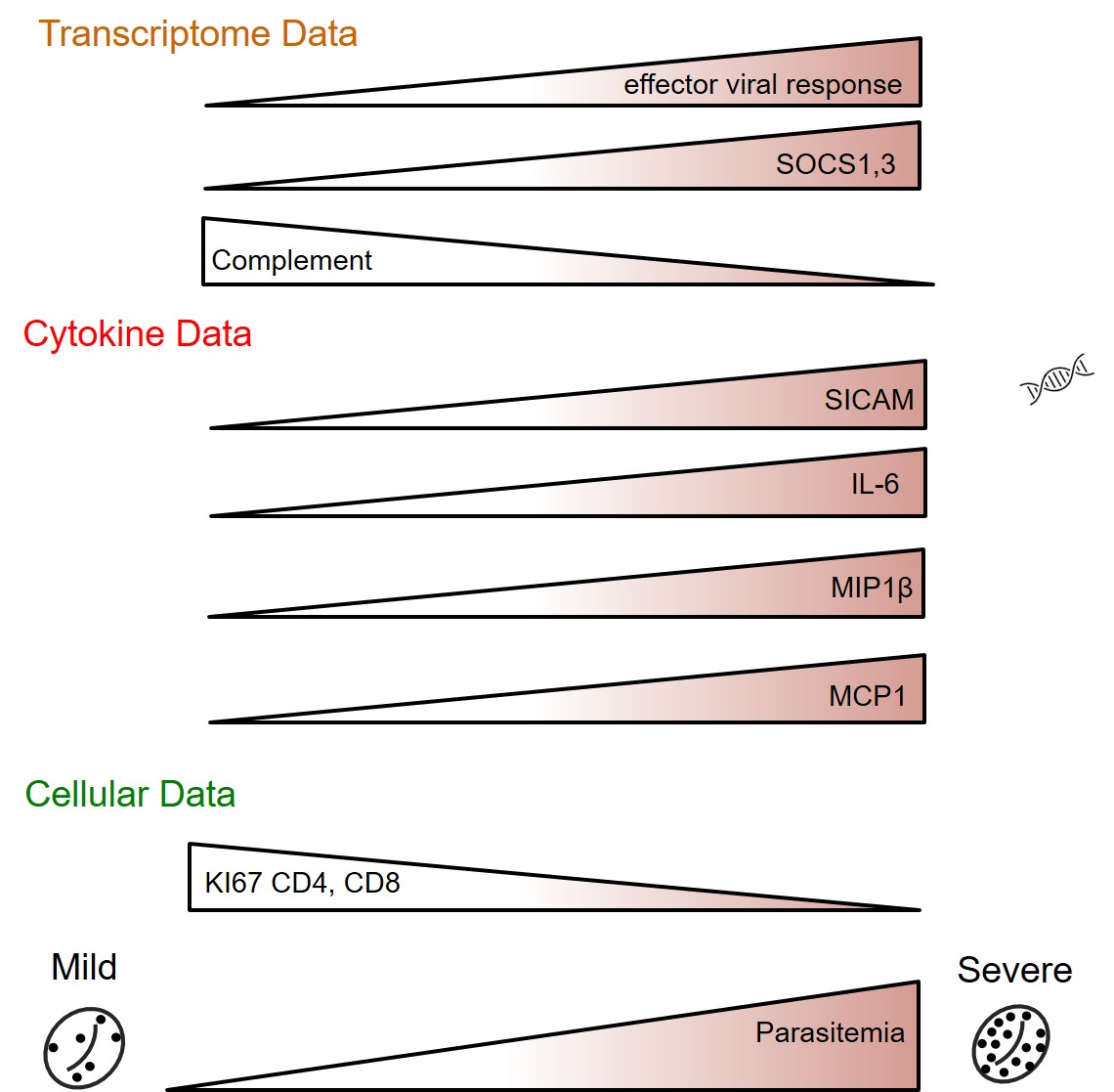}
	\caption{Overview of differential regulation of transcriptome, cytokine, and cellular profiles between subject groups.}
	\label{fig:OV}
\end{figure}

In this study, we analyzed temporal transcriptomic, cellular and cytokine profiles of four \textit{M. mulatta} infected with \textit{P. cynomolgi} to characterize the host response to infection. We leveraged the fact that two of the subjects experienced severe malaria during the primary infection and the other two subjects experienced mild malaria to understand the association of host responses with  \textit{P. cynomolgi} disease severity. The diverse data types allow us to cross-validate transcriptional response with actual protein abundance in relation to cellular populations.

The identification of differentially up-regulated cytokines and transcripts in the severe hosts combined with the downstream pathway analysis demonstrate that severe malaria within our experiment is associated with elevated pro-inflammatory cytokine levels (MIP, IL6, MIG and SICAM1) and the up-regulation of innate immunity and anti-viral responses related gene sets (TLR Pathway, NOD Pathway, and RIG-I Pathway). Additionally, the positive correlations between the abundance of MIP, IL6, MIG and SICAM1 and their respective transcripts suggest that the differential up-regulation of these four cytokines are associated with transcriptional up-regulation. On the other hand, only two transcription factors that activate the production of these cytokines have been identified to be differentially up-regulated. The lack of detection of other transcription factors associated with these cytokines could be attributed to the lack of statistical power due to small sample size, or that the up-regulation of other transcription factors happened prior to our sampling. Additionally, most of the transcriptional regulators that control the expression of the acute phase inflammatory cytokines identified here are controlled by post-translational modifications, and thus, an increase in transcription is not alway observed even though signaling may be occurring. Our characterization of severe malaria caused by malaria infection through elevated IL-6 levels and up-regulation of innate and anti-viral related response have also been observed in studies of \textit{P. falciparum} infection severity \cite{tran2016transcriptomic,mandala2017cytokine,prakash2006clusters,mbengue2016inflammatory}, suggesting likely shared mechanism of recognition of foreign dsRNA and DNA material by the innate immune cells. Correlation analysis of DUGs-S, cell population, and cytokine abundance demonstrates that differential up-regulation of innate immunity and anti-viral responses related gene sets are accompanied by the increase in pro-inflammatory cytokine abundances. The high correlation between cytokine abundance and DUGs-s suggest that the increase in pro-inflammatory cytokines are not independent of transcriptional up-regulation. The negative correlations between DUGs-S and adaptive immune cell types suggest possible innate immune suppression by the adaptive immune system \cite{sun2008inhibition,kim2007adaptive} or cross-talk between innate immune response and adaptive immune response.

On the other hand, the mild hosts are characterized by the differential up-regulation of the complement, coagulation and heme metabolism related genes. Interestingly, the complement and coagulation pathways are differentially down-regulated in the severe hosts. Complement activation has been previously linked to children experiencing severe malaria during \textit{P. falciparum} infection \cite{stoute2003loss,nyakoe2009complement}, however, our result indicate that within the context of \textit{P. cynomolgi} infected \textit{M. mulatta}, elevated levels of complement related transcripts are associated with mild malaria. Correlation analysis of DUGs-M, cell population, and cytokine abundance demonstrates that heme metabolism and complement-related genes are associated with different cell populations and cytokine signals. Complement-related genes are associated with both innate and adaptive immune cells and heme metabolism related genes are associated with parasitemia and inflammation. Interestingly, SICAM1 which is differentially up-regulated in severe hosts at time point 2 shows a significantly higher transcript-cytokine correlation within the severe hosts than that of the mild hosts, which suggest that the abundance of SICAM1 can be further attributed to post-transcriptional activities within the mild hosts such as less cellular damage, inflammation or endothelial activation. Furthermore, IL-1B was observed to be transcriptionally differentially up-regulated in severe hosts, but the large difference of IL-1/IL1RA ratio between the two clinical groups suggest that IL-1 signaling might actually have been hindered in severe hosts.

In summary, we provided transcriptomic, cellular and cytokine evidence associated with severe disease outcome during \textit{P. cynomolgi} infection in \textit{M. mulatta}. We observed that severe malaria is associated with differential up-regulation of innate viral response related genes and pro-inflammatory cytokine, specifically elevated level of IL-6. Mild malaria is associated with up-regulation of the complement pathway strongly associated with monocyte and neutrophil populations. Our analysis provided a molecular and cellular basis for the development of severe malaria during \textit{P. cynomolgi} infection. Larger and more frequent sampling studies are needed to validate our findings and specifically determine the importance of innate immune activation and the role of complement in controlling disease severity.

\section{Material and Methods}
\subsection{Experimental Setup and Data Collection}

A detailed description of the experimental set up and the generation of cytokine and immune profiles can be found in the publication by Joyner \textit{et al.} \cite{joyner2016plasmodium}. The procedure for RNA-seq data collection and the library size normalized data can be found on Gene Omnibus under series number GSE99486. The daily clinical data of the experiment is stored in PlasmoDB using the identifier \textit{E04MalariaClinical}. The cytokine and immune profiling data are deposited on ImmPort under the identifier SDY1015.

\subsection{Bioinformatics Analysis}

Differential expression analysis was carried out using MATLAB's implementation of DESeq algorithm on library size normalized read counts. Enrichment analysis of identified DEG was conducted using MySigDB web service. The gene sets tested for enrichment included canonical pathways, Hallmark pathways, BioCarta pathways and KEGG pathways. A q-value cut-off of 0.05 was used. Clustering of transcriptomic, cellular population and cytokine abundances was done using k-means clustering. The number of clusters used was determined by the silhouette value of different numbers of clusters (1-6).



\begin{thebibliography}{10}
\expandafter\ifx\csname url\endcsname\relax
  \def\url#1{\texttt{#1}}\fi
\expandafter\ifx\csname urlprefix\endcsname\relax\def\urlprefix{URL }\fi
\expandafter\ifx\csname doiprefix\endcsname\relax\def\doiprefix{DOI }\fi
\providecommand{\bibinfo}[2]{#2}
\providecommand{\eprint}[2][]{\url{#2}}

\bibitem{world2016world}
\bibinfo{author}{WHO}.
\newblock \bibinfo{journal}{\bibinfo{title}{World malaria report 2016}}.
\newblock {\emph{\JournalTitle{Geneva: WHO.}}} \textbf{\bibinfo{volume}{13}}
  (\bibinfo{year}{2016}).

\bibitem{mueller2009key}
\bibinfo{author}{Mueller, I.} \emph{et~al.}
\newblock \bibinfo{journal}{\bibinfo{title}{Key gaps in the knowledge of
  plasmodium vivax, a neglected human malaria parasite}}.
\newblock {\emph{\JournalTitle{The Lancet infectious diseases}}}
  \textbf{\bibinfo{volume}{9}}, \bibinfo{pages}{555--566}
  (\bibinfo{year}{2009}).

\bibitem{galinski2013plasmodium}
\bibinfo{author}{Galinski, M.~R.}, \bibinfo{author}{Meyer, E.} \&
  \bibinfo{author}{Barnwell, J.~W.}
\newblock \bibinfo{journal}{\bibinfo{title}{Plasmodium vivax: modern strategies
  to study a persistent parasite’s life cycle}}.
\newblock {\emph{\JournalTitle{Adv Parasitol}}} \textbf{\bibinfo{volume}{81}},
  \bibinfo{pages}{1--26} (\bibinfo{year}{2013}).

\bibitem{noulin20131912}
\bibinfo{author}{Noulin, F.}, \bibinfo{author}{Borlon, C.},
  \bibinfo{author}{Van Den~Abbeele, J.}, \bibinfo{author}{D’Alessandro, U.}
  \& \bibinfo{author}{Erhart, A.}
\newblock \bibinfo{journal}{\bibinfo{title}{1912--2012: a century of research
  on plasmodium vivax in vitro culture}}.
\newblock {\emph{\JournalTitle{Trends in parasitology}}}
  \textbf{\bibinfo{volume}{29}}, \bibinfo{pages}{286--294}
  (\bibinfo{year}{2013}).

\bibitem{deye2012use}
\bibinfo{author}{Deye, G.~A.} \emph{et~al.}
\newblock \bibinfo{journal}{\bibinfo{title}{Use of a rhesus plasmodium
  cynomolgi model to screen for anti-hypnozoite activity of pharmaceutical
  substances}}.
\newblock {\emph{\JournalTitle{The American journal of tropical medicine and
  hygiene}}} \textbf{\bibinfo{volume}{86}}, \bibinfo{pages}{931--935}
  (\bibinfo{year}{2012}).

\bibitem{galinski2008plasmodium}
\bibinfo{author}{Galinski, M.~R.} \& \bibinfo{author}{Barnwell, J.~W.}
\newblock \bibinfo{journal}{\bibinfo{title}{Plasmodium vivax: who cares?}}
\newblock {\emph{\JournalTitle{Malaria Journal}}} \textbf{\bibinfo{volume}{7}},
  \bibinfo{pages}{S9} (\bibinfo{year}{2008}).

\bibitem{tachibana2012plasmodium}
\bibinfo{author}{Tachibana, S.-I.} \emph{et~al.}
\newblock \bibinfo{journal}{\bibinfo{title}{Plasmodium cynomolgi genome
  sequences provide insight into plasmodium vivax and the monkey malaria
  clade}}.
\newblock {\emph{\JournalTitle{Nature genetics}}}
  \textbf{\bibinfo{volume}{44}}, \bibinfo{pages}{1051--1055}
  (\bibinfo{year}{2012}).

\bibitem{waters1993evolutionary}
\bibinfo{author}{Waters, A.~P.}, \bibinfo{author}{Higgins, D.~G.} \&
  \bibinfo{author}{McCutchan, T.}
\newblock \bibinfo{journal}{\bibinfo{title}{Evolutionary relatedness of some
  primate models of plasmodium.}}
\newblock {\emph{\JournalTitle{Molecular biology and evolution}}}
  \textbf{\bibinfo{volume}{10}}, \bibinfo{pages}{914--923}
  (\bibinfo{year}{1993}).

\bibitem{coatneyprimate}
\bibinfo{author}{Coatney, G.}, \bibinfo{author}{Collins, W.},
  \bibinfo{author}{Warren, M.} \& \bibinfo{author}{Contacos, P.}
\newblock \bibinfo{journal}{\bibinfo{title}{The primate malarias. 1971}}.
\newblock {\emph{\JournalTitle{Washington: US Government Printing Office Google
  Scholar}}} .

\bibitem{warren1966biology}
\bibinfo{author}{Warren, M.}, \bibinfo{author}{Skinner, J.} \&
  \bibinfo{author}{Guinn, E.}
\newblock \bibinfo{journal}{\bibinfo{title}{Biology of the simian malarias of
  southeast asia. i. host cell preferences of young trophozoites of four
  species of plasmodium}}.
\newblock {\emph{\JournalTitle{The Journal of parasitology}}}
  \bibinfo{pages}{14--16} (\bibinfo{year}{1966}).

\bibitem{krotoski1982observations}
\bibinfo{author}{Krotoski, W.} \emph{et~al.}
\newblock \bibinfo{journal}{\bibinfo{title}{Observations on early and late
  post-sporozoite tissue stages in primate malaria. ii. the hypnozoite of
  plasmodium cynomolgi bastianellii from 3 to 105 days after infection, and
  detection of 36-to 40-hour pre-erythrocytic forms.}}
\newblock {\emph{\JournalTitle{The American journal of tropical medicine and
  hygiene}}} \textbf{\bibinfo{volume}{31}}, \bibinfo{pages}{211--225}
  (\bibinfo{year}{1982}).

\bibitem{cogswell1992hypnozoite}
\bibinfo{author}{Cogswell, F.~B.}
\newblock \bibinfo{journal}{\bibinfo{title}{The hypnozoite and relapse in
  primate malaria.}}
\newblock {\emph{\JournalTitle{Clinical microbiology reviews}}}
  \textbf{\bibinfo{volume}{5}}, \bibinfo{pages}{26--35} (\bibinfo{year}{1992}).

\bibitem{gonccalves2014parasite}
\bibinfo{author}{Gon{\c{c}}alves, R.~M.}, \bibinfo{author}{Lima, N.~F.} \&
  \bibinfo{author}{Ferreira, M.~U.}
\newblock \bibinfo{journal}{\bibinfo{title}{Parasite virulence, co-infections
  and cytokine balance in malaria}}.
\newblock {\emph{\JournalTitle{Pathogens and global health}}}
  \textbf{\bibinfo{volume}{108}}, \bibinfo{pages}{173--178}
  (\bibinfo{year}{2014}).

\bibitem{prakash2006clusters}
\bibinfo{author}{Prakash, D.} \emph{et~al.}
\newblock \bibinfo{journal}{\bibinfo{title}{Clusters of cytokines determine
  malaria severity in plasmodium falciparum--infected patients from endemic
  areas of central india}}.
\newblock {\emph{\JournalTitle{Journal of Infectious Diseases}}}
  \textbf{\bibinfo{volume}{194}}, \bibinfo{pages}{198--207}
  (\bibinfo{year}{2006}).

\bibitem{tran2016transcriptomic}
\bibinfo{author}{Tran, T.~M.} \emph{et~al.}
\newblock \bibinfo{journal}{\bibinfo{title}{Transcriptomic evidence for
  modulation of host inflammatory responses during febrile plasmodium
  falciparum malaria}}.
\newblock {\emph{\JournalTitle{Scientific Reports}}}
  \textbf{\bibinfo{volume}{6}} (\bibinfo{year}{2016}).

\bibitem{torres2014relationship}
\bibinfo{author}{Torres, K.~J.} \emph{et~al.}
\newblock \bibinfo{journal}{\bibinfo{title}{Relationship of regulatory t cells
  to plasmodium falciparum malaria symptomatology in a hypoendemic region}}.
\newblock {\emph{\JournalTitle{Malaria journal}}}
  \textbf{\bibinfo{volume}{13}}, \bibinfo{pages}{108} (\bibinfo{year}{2014}).

\bibitem{wykes2014malaria}
\bibinfo{author}{Wykes, M.~N.}, \bibinfo{author}{Horne-Debets, J.~M.},
  \bibinfo{author}{Leow, C.-Y.} \& \bibinfo{author}{Karunarathne, D.~S.}
\newblock \bibinfo{journal}{\bibinfo{title}{Malaria drives t cells to
  exhaustion}}.
\newblock {\emph{\JournalTitle{Frontiers in microbiology}}}
  \textbf{\bibinfo{volume}{5}}, \bibinfo{pages}{249} (\bibinfo{year}{2014}).

\bibitem{chavale2012enhanced}
\bibinfo{author}{Chavale, H.}, \bibinfo{author}{Santos-Oliveira, J.~R.},
  \bibinfo{author}{Da-Cruz, A.~M.} \& \bibinfo{author}{Enosse, S.}
\newblock \bibinfo{journal}{\bibinfo{title}{Enhanced t cell activation in
  plasmodium falciparum malaria-infected human immunodeficiency virus-1
  patients from mozambique}}.
\newblock {\emph{\JournalTitle{Mem{\'o}rias do Instituto Oswaldo Cruz}}}
  \textbf{\bibinfo{volume}{107}}, \bibinfo{pages}{985--992}
  (\bibinfo{year}{2012}).

\bibitem{joyner2016plasmodium}
\bibinfo{author}{Joyner, C.} \emph{et~al.}
\newblock \bibinfo{journal}{\bibinfo{title}{Plasmodium cynomolgi infections in
  rhesus macaques display clinical and parasitological features pertinent to
  modelling vivax malaria pathology and relapse infections}}.
\newblock {\emph{\JournalTitle{Malaria Journal}}}
  \textbf{\bibinfo{volume}{15}}, \bibinfo{pages}{451} (\bibinfo{year}{2016}).

\bibitem{anders2010differential}
\bibinfo{author}{Anders, S.} \& \bibinfo{author}{Huber, W.}
\newblock \bibinfo{journal}{\bibinfo{title}{Differential expression analysis
  for sequence count data}}.
\newblock {\emph{\JournalTitle{Genome biology}}} \textbf{\bibinfo{volume}{11}},
  \bibinfo{pages}{R106} (\bibinfo{year}{2010}).

\bibitem{liberzon2011molecular}
\bibinfo{author}{Liberzon, A.} \emph{et~al.}
\newblock \bibinfo{journal}{\bibinfo{title}{Molecular signatures database
  (msigdb) 3.0}}.
\newblock {\emph{\JournalTitle{Bioinformatics}}} \textbf{\bibinfo{volume}{27}},
  \bibinfo{pages}{1739--1740} (\bibinfo{year}{2011}).

\bibitem{han2015trrust}
\bibinfo{author}{Han, H.} \emph{et~al.}
\newblock \bibinfo{journal}{\bibinfo{title}{Trrust: a reference database of
  human transcriptional regulatory interactions}}.
\newblock {\emph{\JournalTitle{Scientific reports}}}
  \textbf{\bibinfo{volume}{5}} (\bibinfo{year}{2015}).

\bibitem{mandala2017cytokine}
\bibinfo{author}{Mandala, W.~L.} \emph{et~al.}
\newblock \bibinfo{journal}{\bibinfo{title}{Cytokine profiles in malawian
  children presenting with uncomplicated malaria, severe malarial anemia, and
  cerebral malaria}}.
\newblock {\emph{\JournalTitle{Clinical and Vaccine Immunology}}}
  \textbf{\bibinfo{volume}{24}}, \bibinfo{pages}{e00533--16}
  (\bibinfo{year}{2017}).

\bibitem{mbengue2016inflammatory}
\bibinfo{author}{Mbengue, B.} \emph{et~al.}
\newblock \bibinfo{journal}{\bibinfo{title}{Inflammatory cytokine and humoral
  responses to plasmodium falciparum glycosylphosphatidylinositols correlates
  with malaria immunity and pathogenesis}}.
\newblock {\emph{\JournalTitle{Immunity, inflammation and disease}}}
  \textbf{\bibinfo{volume}{4}}, \bibinfo{pages}{24--34} (\bibinfo{year}{2016}).

\bibitem{sun2008inhibition}
\bibinfo{author}{Sun, K.} \& \bibinfo{author}{Metzger, D.~W.}
\newblock \bibinfo{journal}{\bibinfo{title}{Inhibition of pulmonary
  antibacterial defense by interferon-$\gamma$ during recovery from influenza
  infection}}.
\newblock {\emph{\JournalTitle{Nature medicine}}}
  \textbf{\bibinfo{volume}{14}}, \bibinfo{pages}{558--564}
  (\bibinfo{year}{2008}).

\bibitem{kim2007adaptive}
\bibinfo{author}{Kim, K.~D.} \emph{et~al.}
\newblock \bibinfo{journal}{\bibinfo{title}{Adaptive immune cells temper
  initial innate responses}}.
\newblock {\emph{\JournalTitle{Nature medicine}}}
  \textbf{\bibinfo{volume}{13}}, \bibinfo{pages}{1248--1252}
  (\bibinfo{year}{2007}).

\bibitem{stoute2003loss}
\bibinfo{author}{Stoute, J.~A.} \emph{et~al.}
\newblock \bibinfo{journal}{\bibinfo{title}{Loss of red blood cell--complement
  regulatory proteins and increased levels of circulating immune complexes are
  associated with severe malarial anemia}}.
\newblock {\emph{\JournalTitle{Journal of Infectious Diseases}}}
  \textbf{\bibinfo{volume}{187}}, \bibinfo{pages}{522--525}
  (\bibinfo{year}{2003}).

\bibitem{nyakoe2009complement}
\bibinfo{author}{Nyakoe, N.~K.}, \bibinfo{author}{Taylor, R.~P.},
  \bibinfo{author}{Makumi, J.~N.} \& \bibinfo{author}{Waitumbi, J.~N.}
\newblock \bibinfo{journal}{\bibinfo{title}{Complement consumption in children
  with plasmodium falciparum malaria}}.
\newblock {\emph{\JournalTitle{Malaria journal}}} \textbf{\bibinfo{volume}{8}},
  \bibinfo{pages}{7} (\bibinfo{year}{2009}).

\end{thebibliography}

\end{document}